\documentstyle[12pt,epsf]{article}
\begin{document}

\title{Microscopics of meson degrees of freedom in nucleons
and mesons in nuclei - what can be seen in the process of quasielastic knockout 
of mesons by high-energy electrons}

\author{V.G. Neudatchin, I.T. Obukhovsky, L.L. Sviridova, N.P.Yudin\thanks{e-mail:yudin@helena.sinp.msu.ru}}
\date{}
\maketitle

\centerline{\small{ D.V. Skobeltsyn Institute of Nuclear Physics, Moscow State University
119899 Moscow, Russia}}
\begin{abstract}
Developed earlier concept of quasielastic knock out of pions from nucleons by
high- energy electrons is propounded as a tool for checking microscopical model
(~$^3P_0$ - fluctuation ) for decay of $N$ to different channels $\pi+ B$ and Preparata
model of nucleus structure.  
\end{abstract}

\section{Inroduction}

1. The quasielastic knock-out (q.e.k.) of protons out of nuclei by high energy electrons  was 
used as mighty tool for investigation of nucleus structure ~\cite{1}. Then the concept of q.e.k. was 
extended ~\cite{2} to the knock-out of electrons out of atoms and molecules and by this  method 
very  important results for the structure of these microobjects were obtained.  For recent 
time in the literature, the notion of quasielastic knock out is often discussed in hadron physics 
~\cite{3}. In general, this term means that in a process almost whole four momentum 
of a  probe is transferred to only one degree of freedom - for instance to nucleon in the 
process of photoproduction of pions on nuclei. 

We have extended this notion to the quasielastic knock out of mesons from nucleon and 
nuclei  and use this notion to clarify some physical problems. 
In general, the set of diagrams gives contributions to the process of photo- or 
electroproduction of pions on nucleons. For example, in Fig. 1 
we show several such 
diagrams (with pseudoscalar interaction). 
\begin{figure}
\epsfverbosetrue
\epsfysize=7cm
\epsfxsize=15cm
\epsfbox{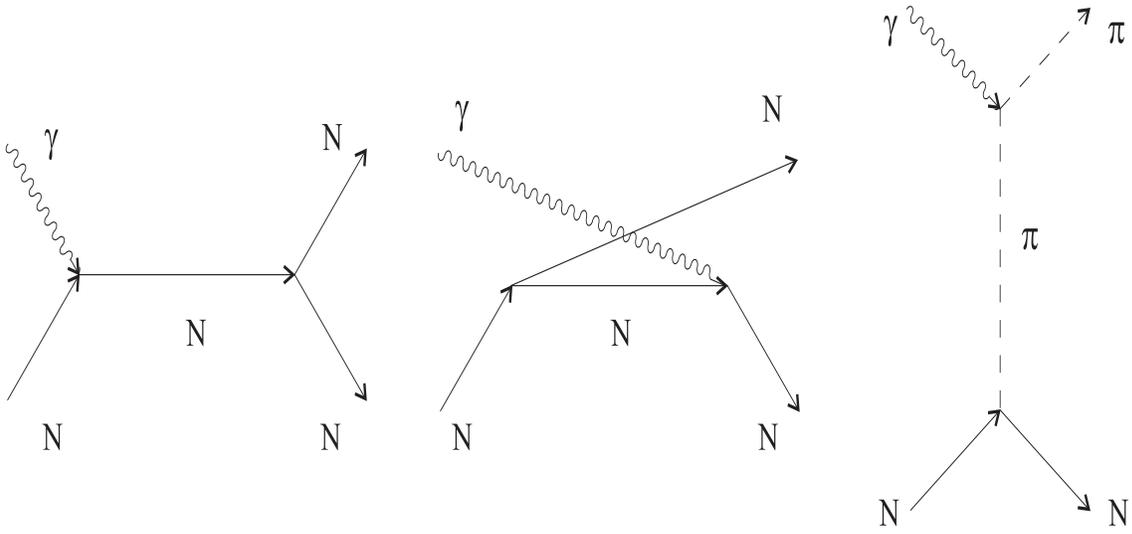}
\caption{The Born diagrams of photo- and electro-production of pions on nucleon.}
\end{figure}
In the case of  photoproduction all of these 
diagrams play very important role. However at sufficiently high 
square momentum transfer $Q^2$ ($Q^2 \ge 1$ (GeV/$c$)$^2$, and total invariant mass of hadrons 
$W \ge$ 2 GeV/$c$ (in order to 
exclude the contributions of resonances)  the situation becomes very simple - for 
longitudinal photons the "pion in flight diagram", that is the diagram with pion pole in 
$t$-channel, becomes dominant. This was shown in our previous papers ~\cite{4}. 
It means that in the longitudinal electroproduction in the mentioned kinematics the  real  
knock-out of pion from nucleon takes place. Thus from experimental data one can extract 
the momentum distribution of pion in nucleon and their spectroscopic factors i.e. the 
number of pions.   This notion opens a new possibility for direct investigation of meson 
structure of nucleon and relevant problems. The similar considerations are also valid for 
electroproduction of pions by transverse virtual photons and we have here the knock-out of 
vector mesons with simultaneous transformation of them into pion (the phenomenon of 
"deexitation" that is similar to the deexitation of alpha - particles in nuclei). It can also 
provide a valuable information on the dynamic of vector mesons in nucleon.

2. Now let's make  several remarks about terms in which we shall discuss these problems. The cross 
section of electroproduction has the following form:

\begin{equation}
\frac{d \sigma(\gamma^*p \to n \pi^+)}{dWdt} \sim \varepsilon \sigma_L +\sigma_T
\end{equation}

where $\varepsilon$ is a usual polarization parameter of the  virtual photon, $t$  - 
momentum square transferred 
to nucleon, $\sigma_L$, $\sigma_T$ - longitudinal and  transverse  cross sections.
Analytically the hadron part of pion in flight diagram of Fig.1 is written as follows:

\begin{equation}
\frac{M(p \to n \pi^+)}{k^2-m_\pi^2}=\frac{M(p \to n \pi^+)}{k_0-e_k} \cdot \frac{1}{k_0+e_k}=
<n|a_{\pi^+}|p> \frac{1}{k_0+e_k}
\end{equation}
              
where  $a_{\pi^+}$ is an operator of virtual pion absorption, $k_0$ - zero component of transferred 
from nucleon to photon four vector, $e_k$  - energy of the virtual pion with momentum $k$ in the 
rest frame of proton, $p$ and $n$ - proton and neutron, $M(p \to n \pi^+)$ - invariant amplitude of 
transformation  $M(p \to n \pi^+)$. It is important to point out that this  equation  takes into account 
simultaneously the diagrams going "forward and backward" in time, and that we work in 
the lab system. The quantity $<n|a_{\pi^+}|p>$ defines the momentum distribution of pions in 
proton and we call it "wave function"  and designate by $\Psi_p^{n \pi}$. This wave 
function is normalized by the condition:

\begin{equation}
\int d \tau |\Psi_p^{n \pi}(k)|^2=S_p^{n \pi}
\end{equation}

where  $d \tau = d^3k/((4 \pi)^3E_pE_n)$, $E_{p,n}$ - energies of proton and neutron, $S_p^{n \pi}$ is a 
spectroscopic factor of pion in nucleon. It determines the number of pions participating in 
production of the channel. The sum of the spectroscopic factors over all final channels 
will  give the total number of positive pions in nucleon.

As a result, we get that the longitudinal cross section of pion electroproduction has the 
following form:

\begin{equation}
\frac{d \sigma_L(\gamma^*p \to n \pi^+)}{dt} \sim |\Psi_p^{n \pi}(k)|^2 F_{\gamma \pi \pi}^2(Q^2)
\end{equation}

where $t = k^2$  and $ F_{\gamma \pi \pi}^2(Q^2)$ - electromagnetic formfactor of pion, $Q^2$ - 
"mass" of the virtual photon. 

\section{Check of $^3P_0$ fluctuation model
}

It is well known, that the hadrons are effective degrees of freedom and the hadron 
dynamics should be a phenomenological one. The important ingredient of this 
phenomenology is the coupling constants of meson with baryons. Now when an "attack" on  
hadrons world is wide underway the number of constant becomes big and it is necessary to 
search for notions, that can  unite the description of hadrons. It is naturally, that the search for 
unification should be based on quark-gluon degrees of freedom. One way to realize this 
unification is to introduce the coupling between quarks and mesons. Then the coupling with 
baryons comes out as a result of some averaging over quark structure of baryons. Partially 
this was done in the paper ~\cite{5}. But this approach entails also  many parameters. It is proved 
that an old model of $^3P_0$  fluctuation has definite advantages. On diagram level this 
model is presented in Fig.2.
\vspace*{-3.cm}
\begin{figure}[h]
\epsfverbosetrue
\epsfysize=7cm
\epsfxsize=10cm
\begin{center}
\epsfbox{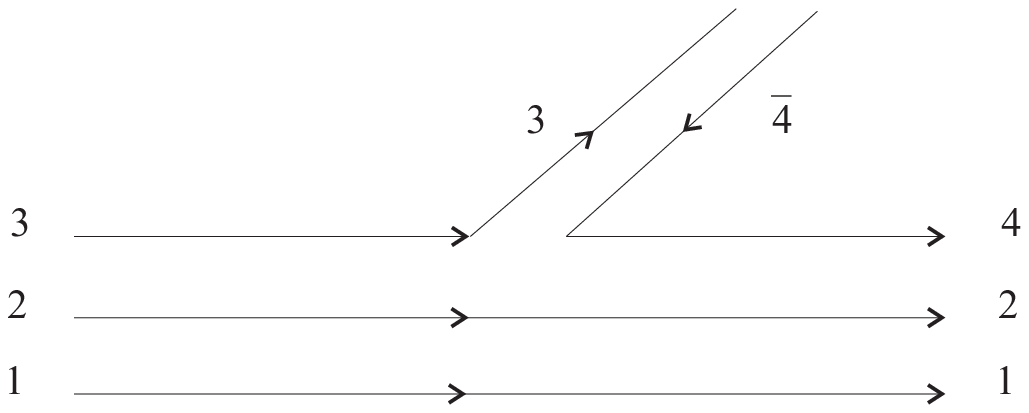}
\end{center}
\caption{The diagram of vacuum production of the flavour SU3-invariant  
quark-antiquark 
 pair.}
\end{figure}

The advantage of this approach based on the fact that the five quarks in the final system can be 
transformed into hadron by many different ways. This permits one on the same footing to 
consider and therefore to  relate the whole set of meson-baryons coupling constants, for 
instance pion-baryon, kaon-baryon, vector meson-baryon and so on.
The formal aspects of this rearrangement were considered in our papers ~\cite{6,7} and will not 
be discussed here. Calculated momentum distributions and formfactors of pions and kaons 
in nucleon are shown in Fig.3,4. 

\begin{figure}[h]
\parbox{7.cm}{
\epsfverbosetrue
\epsfysize=7cm
\epsfxsize=7cm
\epsfbox{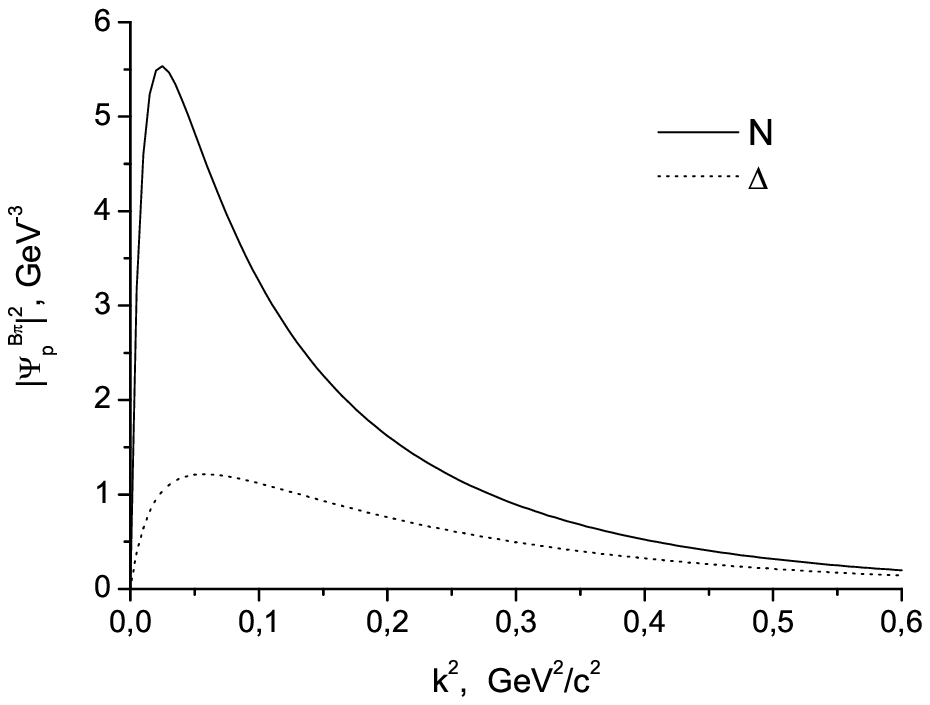}
\caption{The momentum distributions of pions $|\Psi_p^{B \pi}|^2$
in different channels ($\pi B=\pi N, \pi \Delta$).}
}
\hfill
\parbox{7.cm}{
\epsfverbosetrue
\epsfysize=7cm
\epsfxsize=7cm
\epsfbox{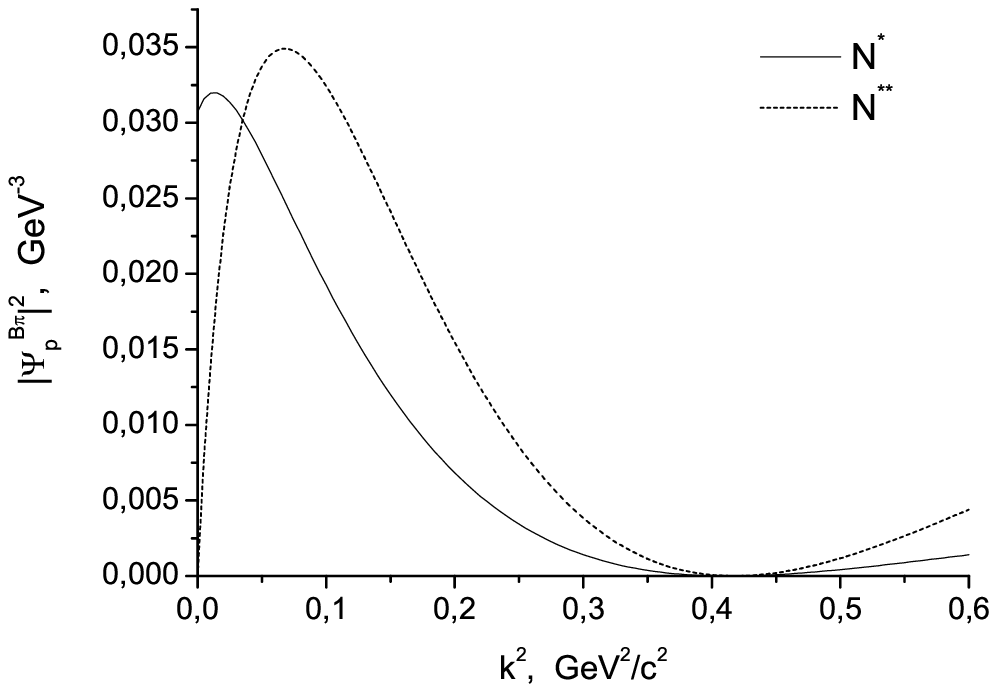}
\caption{The momentum distributions of pions $|\Psi_p^{B \pi}|^2$
in different channels ($B=N^*, N^{**}). N^* - N(1535), N^{**} - N(1440).$}
}
\end{figure}

One can see that discussed momentum distributions are 
very interesting and they demand further investigation. 

In conclusion we would like to remark that $^3P_0$-fluctuation 
is probably closely related with scalar quark condensate. This is one more 
reason to investigate  this model.

\section{Mesons in nuclei}

Over past years, the problem of pions in nuclei was  discussed in literature ~\cite{8} many times. 
For instance, a very well known paper has a title "Where is the pions in nuclei" ~\cite{9}.
We can nott dwell on the details of this very interesting questions and confine ourselves with the 
statement that now the problem is partially clear - in any case, the pions in nuclei predicted 
by RPA approach were detected and  sharp contradiction between prediction of the theory and 
(p,n) experiment, that was a base for the title of paper ~\cite{9}, was probably removed. In a sense,
one really sees the pions in nuclei injected there by RPA correlations ~\cite{10}. New nontrivial 
aspects of the nuclear  pion problem  were considered  in the paper ~\cite{11}.

Nevertheless the problem of mesons in nuclei is far from its full solution and probably 
that we will meet here the unexpected and nontrivial phenomena.

The first modern nontrivial notion of meson's injection into nuclei was propounded by 
Migdal ~\cite{12}. Now it seems clear that his notion of pion condensate in nuclei is not realized. 
But there exists another concept of pion condensate propounded by Preparata ~\cite{13}. 
Preparata's idea is closely related to the effect of Dicke ~\cite{14} that is known in atomic physics 
for several tens of years.

The point is: let's suppose that one has a system of $N$ exited atoms. Then as long as  
radioactive transitions take place the probability of transitions becomes bigger and bigger 
and in the "middle" - i.e. when about one half of exited atoms is left - the probability 
becomes as much as $N^2/4$. At large $N$ the increase of probability becomes fantastic!
Preparata has assumed that in nuclei something similar can take place if one takes into account 
$\Delta$-resonance. In this case, one has 
$\Delta \to$ excited atoms, $N \to$ atom in the ground state, pion $\to$   
photon. 
According to an estimate by Preparata if one takes into account  the existence of 
such transitions the increasing of self-energy of whole nucleus will result in gain of energy 
that amounts 60 MeV per one nucleon! Thus,  a heavy nucleus can consist of  
three coherently vibrating subsystems $\Delta$-nucleons, $\Delta$-resonances and pions. With the usual 
way of investigation it is not simple task to observe such a substructures. We suggest to use 
for this purpose the quasielastic knock out of pions. In Fig.5,\\ 
\vspace*{-1cm}
\begin{figure}[h]
\epsfverbosetrue
\epsfysize=7cm
\epsfxsize=10cm
\begin{center}
\epsfbox{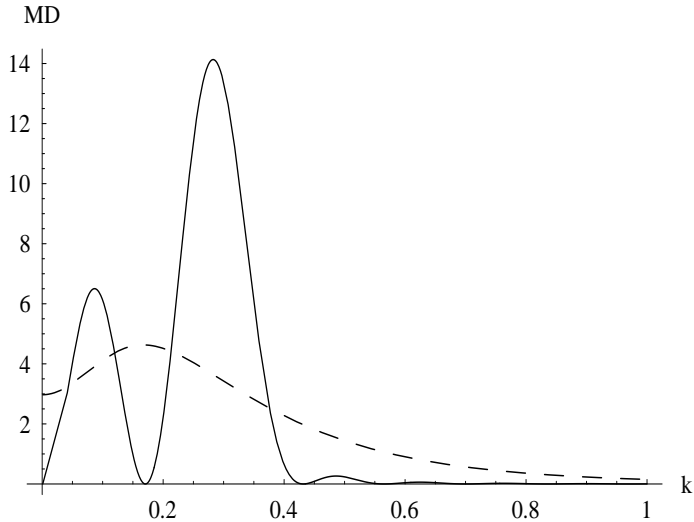}
\end{center}
\caption{Momentum distributions of pions in nuclei: solid line - 
the MD $|\Psi_p^{n \pi}(k)|^2$, (GeV/$c$)$^{-3}$ versus $|k|$, GeV/$c$ in Preparata model;
dashed line - the MD in the model of independent nucleons.}
\end{figure}

\noindent it is shown the momentum distribution of 
pions that are injected into nuclei by Preparata mechanism in comparison with trivial
injection by independent nucleons. We see that the momentum distributions of pions, 
injected by usual mechanisms (independent nucleons) and Preparata mechanism are 
drastically different and measurement of pion momentum distribution helps to prove whether 
there exists a collective strengthening of pions in nuclei.



\begin{thebibliography}{99}

\bibitem{1}
D.G. Ireland and G.Van der Steenhoven Phys.Rev. C49,2182(1994); D.Debruyne, 
      J.Ryckebush et al Phys.Rev.C62, 024611 (2000).
      
\bibitem{2} V.G.Neudatchin, Yu.V. Popov and Yu.F. Smirnov Usp.Fiz.Nauk 169,1111(1999); 
      E.Weigold and McCarthy Electron Momentum Spectroscopy, Plenum, New York,1999.

\bibitem{3} X.Li and L.E.Wright, C.Bennhold Phys.Rev.C48,816 (1993).

\bibitem{4} V.G.Neudatchin, N.P.Yudin and L.L.Sviridova Yad.Fiz. 64,1680 (2001)

\bibitem{5} G.Brown and D.Riska Nucl.Phys.A679 (2001),577

\bibitem{6} I.T.Obukhovsky, V.G.Neudatchin, L.L. Sviridova, N.P. Yudin Yad.Fiz. 66(2003)

\bibitem{7} I.T.Obukhovsky, V.G.Neudatchin, L.L.Sviridova Yad.Fiz. to be published 

\bibitem{8} D.S.Koltun   Phys.Rev. C57,1210 (1997)

\bibitem{9} G.E.Brown, M,Bubala,Zi.Bang, J.Wambach Nucl.Phys.A593(1995)293.

\bibitem{10} H.Sakai Nucl.Phys.A690(2001)66c.

\bibitem{11} G.Chanfray,M.Ericson and P.A.M.Guichon Phys.Rev. C63(2001)055202.

\bibitem{12} A.B. Migdal Rev.Mod.Phys. 50,107(1978).

\bibitem{13} G.Preparata Nuovo Cimento A103(1990)1213.R.Alzetta,G. Liberti and G.Preparata 
       Nuovo Cimento A112(1999)1609 

\bibitem{14} A.V. Andreev,V.I. Emel'yanov and L.L. Il'inskii  Coherent Phenomena in Optics 
        (Nauka,Moscow,1988).

\end{thebibliography}
\end{document}